\documentclass[aps,twocolumn,prl,tightenlines,floatfix,showpacs]{revtex4}

\usepackage[dvips]{graphicx}
\usepackage[english]{babel}
\usepackage{amsmath}
\usepackage{amssymb}
\usepackage{times}
\usepackage{multirow}

\begin{document}

\newcommand{\D}{\mathrm{D}}
\newcommand{\p}{\partial}
\newcommand{\Tr}{\mathrm{Tr}}

\title{Establishing the Presence of Coherence in Atomic Fermi Superfluids: 
Spin-Flip and Spin-Preserving Bragg Scattering at Finite Temperatures}

\author{Hao Guo $^{1}$, Chih-Chun Chien$^{2}$ and K. Levin$^{1}$}

\affiliation{$^1$James Franck Institute and Department of Physics,
University of Chicago, Chicago, Illinois 60637, USA}

\affiliation{$^2$Theoretical Division, Los Alamos National Laboratory, MS B213, Los Alamos, NM 87545, USA}

\date{\today}

\begin{abstract}
We show how the difference between the finite temperature
$T$ structure factors, called $S_-$,  associated
with spin and density, can be used as a indication of superfluidity
in ultracold Fermi gases.
This observation can be exploited in two photon Bragg scattering experiments
on gases which undergo BCS- Bose Einstein condensation 
crossover. Essential to our calculations is a proper incorporation of
spin and particle number conservation laws which lead to compatibility at
general $T$ with
two $f$-sum rules. 
Because it is applicable to general scattering lengths,
a measurement of $S_-$ can be
a useful, direct approach for establishing where superfluidity occurs.
\end{abstract}


\maketitle

One of the most intriguing aspects of the ultracold Fermi gases
in the strongly interacting, attractive regime is the difficulty
they pose for establishing superfluidity. The absence of
indications from the density profiles,
along with the presence of a
normal state excitation gap are
all responsible for this difficulty.
Most experiments
\cite{Jin4,Ketterle3,KetterleV,Inada}
which determine whether a given state is above or below the transition at
$T_c$,
make use of a fast sweep of the magnetic field into
the low field Bose Einstein condensation (BEC) limit, where
the superfluid signatures are more straightforward.
Thermodynamical properties
do not establish whether the phase is ordered or not, although they
can determine where $T_c$ lies \cite{ThermoScience,SalomonFL}.
It would be particularly useful then to find a
neutral fluid counterpart to the Meissner experiment for Fermi gases
with attractive interactions.
More specifically, we want to
find a property which
reflects the
superfluid order parameter such that it is non-zero below $T_c$ and
vanishes above.
Of particular interest is Bragg spectroscopy \cite{valePRL08}
which is capable of detecting true superfluid coherence.
This is to be contrasted with radio frequency spectroscopy \cite{ourRFreview09}
which is most suited for establishing pairing.

It is the purpose of this paper to
address this goal by presenting a theory of the
spin preserving and spin flip dynamical responses at general temperatures
in a homogenous gas. Both
of these can, in principle,
be measured via two photon Bragg spectroscopy.
Our emphasis here is on the contrasting physics of spin density and particle
density which
we demonstrate allows a separation of coherent order from
pairing effects. While we focus on unitary
gases, we also present results associated with both the BCS and
the BEC sides of resonance. Indeed, there has been considerable
theoretical interest in combined studies of spin
and particle density Bragg scattering at zero
temperature in BCS-BEC crossover theory
\cite{Combescot06,ZwergerPRL04}
as well as of spin-flip Bragg experiments above $T_c$
\cite{BaymPRA06}.

Two aspects are essential to a proper theory of Bragg
scattering. (i) At all temperatures the identities 
based on conservation laws for particle number and spin need to be satisfied:
\begin{equation} 
\int^{\infty}_{-\infty}d\omega\omega S_{C,S}(\omega,\mathbf{q})
=n_F\frac{\mathbf{q}^2}{m},\label{fs1}
\end{equation}
\begin{equation}
\lim_{\mathbf{q}\rightarrow\mathbf{0}}S_{C,S}(\omega,\mathbf{q})=0, \label{fs2}
\end{equation}
where $S_{C,S}(\omega,\mathbf{q})$ are the dynamical structure factors
for the particle density (labelled $C$) and the spin (labelled $S$).
Here $n_F$ is the total density of particles.  The first equation is
the well known $f$-sum rule in the form associated with the
generalized BCS Hamiltonian.  In past work \cite{BaymPRA06} sum rules
were found to be difficult to implement in the presence of
pairing correlations.   (ii) Below
$T_c$, collective mode physics in the charge channel is responsible for
Eq.~(\ref{fs2}), not Pauli blocking as is sometimes
claimed; this ensures
current conservation and restores the $U(1)$ gauge symmetry.

We define
\begin{equation}
S_{\pm}(\omega,\mathbf{q}) \equiv \frac{1}{2}[S_{C} (\omega,\mathbf{q}) \pm
S_{S} (\omega,\mathbf{q})]
\label{eq:3}
\end{equation}
From Eq.~(\ref{fs1}) an important additional sum rule then follows
\begin{eqnarray} \label{SS-}
\int^{\infty}_{-\infty}d\omega\omega S_-(\omega,\mathbf{q})=0.
\end{eqnarray}
The quantity $S_{-}(\omega, \mathbf{q})$ represents a measure of
``spin density and particle density separation''.  
In the literature \cite{Combescot06} the quantity
$S_{\uparrow \downarrow}/2$ is defined in the same way as $S_{-}$.
Importantly, $S_{\uparrow \downarrow}$, has been argued
\cite{Drummond10,Combescot06} to be proportional to the density
density correlation function $\langle\rho_{\uparrow}(r)
\rho_{\downarrow}(r)\rangle$.  The sum rule of Eq.(\ref{SS-})
indicates that $S_{\uparrow \downarrow}$ times frequency integrates to
zero.  In strict BCS theory this sum rule is satisfied because 
$S_{-}(\omega, \mathbf{q})$ (or equivalently $S_{\uparrow
\downarrow}(\omega, \mathbf{q})$) has negative weight below $T_c$ and
is identically zero above $T_c$.  
In this paper
we show that the behavior found in strict BCS
is general.
On this basis,  we argue that
\textit{the association of the density-density correlation function
(with opposite spin) and the quantity $S_{\uparrow \downarrow}(\omega,
\mathbf{q})$ (or equivalently $S_{-}(\omega, \mathbf{q})$) is very
problematic}. Collective mode effects are, in part, the reason that
naive linear response theory (relating
 $\langle\rho_{\uparrow}(r)
\rho_{\downarrow}(r)\rangle$ to
$S_{\uparrow \downarrow}$) fails below $T_c$.

The theoretical approach used here is based on the BCS-Leggett ground
state wavefunction, which has been shown \cite{Combescot06} to give a
reasonable description of the $T=0$ spin and density dynamical
structure factors as computed using quantum Monte Carlo
simulations. Here we generalize this treatment of BCS-BEC crossover to
finite $T$ and must choose the appropriate diagram set to respect the
conservation laws in Eqs.(\ref{fs1}) and (\ref{fs2}). The present
approach is most well suited to address these sensitive issues which
arise in gauge invariant electrodynamics; here in contrast to other
crossover schemes \cite{firstordertransitionpapers} the superfluid
density is well behaved and monotonic in $T$.  We build on the analysis
outlined in Ref.~\cite{Kosztin00} which focused principally on
collective mode effects.  In this extension of the ground state to
finite $T$, the pairing gap $\Delta(T)$ has two contributions
associated with condensed ($\textrm{sc}$) and non-condensed
($\textrm{pg}$) pairs such that $\Delta^2(T) =
\Delta_{\textrm{sc}}^2(T) + \Delta_{\textrm{pg}}^2(T)$.  The latter
vanishes at $T=0$ and the former vanishes at $T_c$ and above.

\begin{figure}[tb]
\centerline{\includegraphics[clip,width=3.0in]{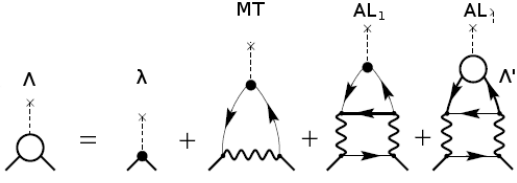}}
\caption{The diagrams contributing to the full electromagnetic vertex
  $\Lambda$. Here the wiggly lines represent
  the $T$-matrix $t_{pg}$ and the dashed line represent the singular
  ``condensate propagator'' $t_{sc}$. In AL$_2$,
  $\Lambda^\prime$ is the full gauge invariant EM vertex.}
\label{fig:full_vertex}
\end{figure}

For spin-preserving and spin flip Bragg processes \cite{wordsBragg}
the external field Hamiltonian contains contributions of the form
$\int d^3\mathbf{r}[\lambda_{\sigma\sigma^{\prime}}^{C,S}
\Psi_{\sigma}^{\dagger} \Psi_{\sigma^{\prime}}+\textrm{h.c.}]$ where
$\lambda^C_{\sigma\sigma^{\prime}}\propto
\delta_{\sigma\sigma^{\prime}} $ and $
\lambda^{S}_{\sigma\sigma^{\prime}} \propto
\delta_{\sigma\bar{\sigma}^{\prime}} $. Here $\Psi^{\dagger}$($\Psi$)
is the fermionic creation (annihilation) operator,
$\sigma=\uparrow,\downarrow$, $\uparrow=-\downarrow$ and
$\bar{\sigma}=-\sigma$.  While there would seem to be a similarity in
these two channels we stress that there is also a physical
difference. In the superfluid phase, due to coherent singlet pairing,
the response functions of spin and particle densities are different,
whereas one may anticipate that they are the same in the normal
phase. Indeed, linear response theory based on the above Hamiltonian
is incomplete; it will not lead to a consistent treatment of the
density preserving response, because it ignores collective mode
effects below $T_c$.  For the unperturbed Hamiltonian of the BCS type,
and for singlet pairing, spin is conserved. 
However, because the spin degrees of freedom are not
associated with a spontaneous symmetry breaking, they do not lead to
collective phase modes upon condensation.
\begin{figure}[tb]
\centerline{\includegraphics[clip,width=3.0in]{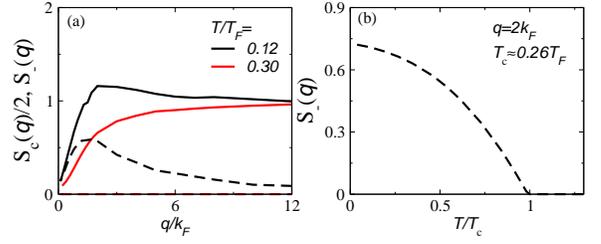}}
\caption{(Color online) (a) Static structure factor (in units of $n_F/2$) at unitarity in the 
density channel $S_c(\mathbf{q})/2$ (solid lines) as a function of $q$
for two temperatures and $S_-(\mathbf{q})$ (dashed lines).  (b)
Temperature dependence of $S_-(\mathbf{q})$ showing that it behaves
like an order parameter, as in BCS theory and for all $k_Fa$. Here $k_F$, $\epsilon_{F}$, and $T_F$ are the Fermi momentum, Fermi energy, and Fermi temperature of a non-interacting Fermi gas with the same particle density.}
\label{fig:StaticS}
\end{figure}

In the density channel, 
linear response theory is more complex; we introduce 
a four-vector formalism with 
``electromagnetic'' (EM) field 
$A_{\mu}$ where $\mu=0,1,2,3$. 
The induced EM current $J_{\mu}$ is given by 
$J^{\mu}(Q)=K^{\mu\nu}(Q)A_{\nu}(Q)$, where, $Q=q_{\mu}=(\omega,\mathbf{q})$ is the 
four-momentum. and the EM response kernel, $K^{\mu\nu}$, can be written as 
$K^{\mu\nu}(Q)=K^{\mu\nu}_0(Q)+\delta K^{\mu\nu}(Q)$, where $\delta K^{\mu\nu}$ is 
the correction due to collective mode effects
\cite{Kosztin00}. Here $\delta K^{\mu\nu}$, which is essential
for charge conservation ($q_{\mu}K^{\mu\nu}(Q)=0$), has a vanishing
denominator \cite{Kosztin00} associated with the order parameter
collective mode dispersion.  In this letter, we are interested in the
$00$-component of the EM kernel, which corresponds to the gauge
invariant density-density response function
$\chi_{\rho\rho}(\omega,\mathbf{q})\equiv K^{00}(\omega,\mathbf{q})$.
This, in turn, is related to the dynamic structure factor in the
density channel, given by
$S_{C}(\omega,\mathbf{q})=-\frac{1}{\pi}\textrm{coth}
(\frac{\omega}{2T})\textrm{Im}\chi_{\rho\rho}(\omega,\mathbf{q})$,
where $\chi_{\rho\rho}(\omega,\mathbf{q})$ represents the properly
gauge invariant form of the density-density response function. In a similar way, the dynamic spin structure factor is $ 
S_{S}(\omega,\mathbf{q})=-\frac{1}{\pi}\textrm{coth}
(\frac{\omega}{2T})\textrm{Im}\chi_{SS}(\omega,\mathbf{q})$, where
$\chi_{SS}(\omega,\mathbf{q})$
is the spin-spin response function.

Now we turn to finite temperatures where there is little prior work
addressing Bragg scattering in the crossover scenario. Charge
conservation and gauge invariance which can be explicitly shown to
hold at the BCS level must be respected
\cite{Kosztin00} even in the presence of pairing correlations.
The best way to insure these is to establish that the $f$-sum rules
Eq.~(\ref{fs1}) and the related Eq.~(\ref{fs2}) are satisfied
\cite{wordscollect}.  For
our BCS-Leggett ground state, extended to $T \neq 0$, the diagrammatic
corrections to the EM vertex have been discussed \cite{Kosztin00} and
are indicated in Fig.\ref{fig:full_vertex}. There are two types of
contributions shown as Maki-Thompson (MT) and Aslamazov-Larkin
(AL$_{1,2}$) diagrams.

The ``bare'' part of the EM 
response kernel is given by $K^{\mu\nu}_0(Q)-\frac{n_F}{m}g^{\mu\nu}(1-g^{\mu0})=$ 
\begin{equation} 2\sum_{K}\lambda^{\mu}(K,K-Q)G(K)G(K-Q)\Lambda^{\nu}(K,K-Q),\\ 
\end{equation} 
where the bare and full EM vertex are
$\lambda_{\mu}(K,K+Q)=(1,\frac{1}{m}(\mathbf{k}-\frac{\mathbf{q}}{2}))$
and
$\Lambda_{\mu}=\lambda_{\mu}+\delta\Lambda_{\textrm{sc}\mu}+\delta\Lambda_{\textrm{pg}\mu}$
respectively. Here $m$ is the particle mass and $g^{\mu\nu}$ is the
metric tensor. Note that the full vertex $\Lambda$ does not include
collective mode physics and, thus, is not gauge invariant below $T_c$,
while the vertex $\Lambda^{\prime}$ contained in the AL$_2$ diagram is
gauge invariant.  Importantly, consistency with the Ward identities
will lead a cancellation
\cite{Kosztin00}
between the MT and AL terms
$\frac{1}{2}(\textrm{AL}_1+\textrm{AL}_2)+\textrm{MT}^C_{\textrm{pg}}=0$. Throughout
we use the superscript ``C'' to refer to the density response and the
counterpart ``S'' for the spin counterpart.

\begin{table}
\begin{tabular}{|c|c|c|c|c|}
\hline
$\frac{1}{k_Fa}$ & $\frac{T}{T_F}$ & $\int^{\omega_M}_{-\omega_M}d\omega\omega S_{C}(\omega,\mathbf{q})$ & $\int^{\omega_M}_{-\omega_M}d\omega\omega S_S(\omega,\mathbf{q})$ & $\frac{\omega_M}{k_F}$ \\
\hline
\multirow{2}{*}{-1.0} & 0.06 & 1.000 & 0.991 & \multirow {2}{*}{$\sim$ 20} \\\cline{2-4}
& 0.12 & 0.999 & 0.999 &\\
\hline
\multirow{2}{*}{-0.5} & 0.10 & 0.996 & 0.972 & \multirow {2}{*}{$\sim$ 50}\\\cline{2-4}
& 0.22 & 1.020 & 1.020 & \\
\hline
\multirow{2}{*}{0.0} & 0.12 & 0.976 & 0.941 & \multirow {2}{*}{$\sim$ 100}\\\cline{2-4}
& 0.28 & 0.964 & 0.964 &\\
\hline
\multirow{2}{*}{1.0} & 0.05 & 0.902 & 0.909 & \multirow {2}{*}{$\sim$ 500}\\\cline{2-4}
& 0.25 & 0.910 & 0.910 &\\
\hline
\end{tabular}
\caption{$f$-sum rule tests for density and spin structure factors in BCS-BEC crossover.
$\omega_M$ is the maximum frequency. The expected value 
is $1.0$ in units of $n_F\epsilon_F/2$.}
\label{tab1}
\end{table}
After imposing the Ward identities \cite{Kosztin00}, 
the dynamical structure factor for particle density can be 
written as the sum of five terms 
\begin{eqnarray}\label{Scf} 
S_C=S_{C0}+S_{\textrm{MT}^C_{\textrm{sc}}}+S_{\textrm{MT}^C_{\textrm{pg}}}+S_{\textrm{AL}}+S_{\textrm{coll}}, 
\end{eqnarray} 
where the first term on the right-hand-side denotes the ``bare'' term from 
$K^{00}_0$. The remaining terms denote the 
corrections from the MT, and two AL 
diagrams and from the collective modes.

In the spin channel, the spin-``magnetic field'' interaction also
contains effects associated with pairing fluctuations, but there is a
significant difference. Here the AL diagrams are not included. This
assertion can be verified by establishing numerical consistency with
the $f$-sum rule (\ref{fs1}). It is also, then, consistent with the
equality in Eq.~(\ref{fs2}).  As a result, the full vertex is given by 
$\Lambda_S=\lambda_S+\delta\Lambda_{\textrm{MT}^S_{\textrm{sc}}}+\delta\Lambda_{\textrm{MT}^S
_{\textrm{pg}}}$.  Here the bare spin-magnetic field interaction
vertex is $\lambda_S(P,P+Q)=1$. Diagrammatically these contributions
are represented by the first two terms on the right hand side in
Fig.~\ref{fig:full_vertex}. The full spin response function is given by
$\chi_{SS}(Q)=$ \begin{equation}
2\sum_{K}\lambda_S(K,K-Q)G(K)G(K-Q)\Lambda_S(K,K-Q).\\ \end{equation}
We arrive at a final
compact expression for the spin structure factor 
\begin{eqnarray}\label{Ssf}
S_S=S_{S0}+S_{\textrm{MT}^S_{\textrm{sc}}}+S_{\textrm{MT}^S_{\textrm{pg}}}.
\end{eqnarray}

\begin{figure}[tb] 
\centerline{\includegraphics[clip,width=3.0in]{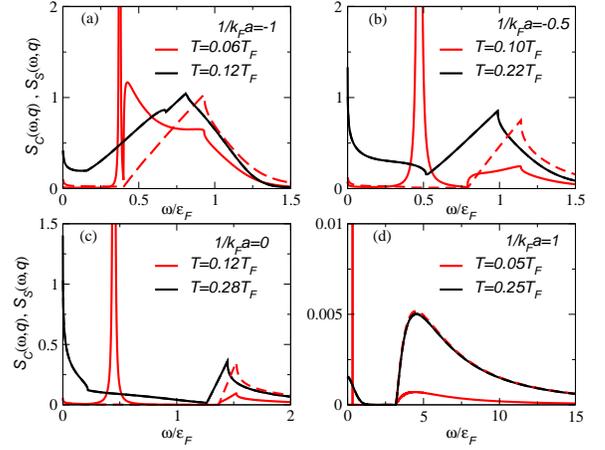}} 
\caption{(Color online) Dynamic 
density structure factor (solid lines) and spin structure factor
(dashed lines) in BCS-BEC crossover for $q=0.5k_F$. The black solid
and dashed lines which show the behavior slightly above $T_c$,
coincide with each other since $S_C=S_S$. The red lines indicate the
behavior below $T_c$ where $S_C \neq S_S$.}
\label{fig:DDS} 
\end{figure}

We now evaluate $S_{-}(\omega, \mathbf{q})$.  It can be shown that
$S_{C0}=S_{S0}$. Moreover, the MT terms in the spin and density
channel enter with reversed signs, as has been noted previously
\cite{ZwergerPRL04}:
$S_{\textrm{MT}^C_{\textrm{sc}}}=-S_{\textrm{MT}^S_{\textrm{sc}}}$ and
$S_{\textrm{MT}^C_{\textrm{pg}}}=-S_{\textrm{MT}^S_{\textrm{pg}}}$. Together
with the cancellation between MT and AL terms, we arrive at a very
general form 
for $S_{-}$, defined in Eq.({\ref{eq:3}}).
\begin{eqnarray}\label{Spm}
S_-=S_{\textrm{MT}^C_{\textrm{sc}}}+\frac{1}{2}S_{\textrm{coll}}.
\end{eqnarray}
\textit{Importantly,
$S_{-}(\omega, \mathbf{q})$
depends only on the order parameter}; it
vanishes above $T_c$, as is trivially
consistent with the sum rule
in Eq.~(\ref{SS-}).
For the strict BCS case, where the same behavior obtains, one
can clearly see that this is not a consequence of the vanishing
of the gap, but rather of true phase coherence.
Above $T_c$ the same diagrams leading to
the vanishing of the superfluid density \cite{Kosztin00}
are responsible
for the vanishing of $S_{-}$.

This discussion has been general; now we substitute our specific
form for the self energy \cite{Kosztin00}
to obtain
\begin{eqnarray}\label{Chif}
& &K^{00}_0(\omega,\mathbf{q})\textrm{ or }\chi_{SS}(\omega,\mathbf{q})=\sum_{\mathbf{p}}\Big[(1-f_+-f_-)\times\nonumber\\
& &\frac{E_++E_-}{E_+E_-}\frac{E_+E_--\xi_+\xi_-+\textrm{sgn}\Delta^2_{\textrm{sc}}-\Delta^2_{\textrm{pg}}}{\omega^2-(E_++E_-)^2}-(f_+-f_-)\nonumber\\
& &\times\frac{E_+-E_-}{E_+E_-}\frac{E_+E_-+\xi_+\xi_--\textrm{sgn}\Delta^2_{\textrm{sc}}+\Delta^2_{\textrm{pg}}}{\omega^2-(E_+-E_-)^2}\Big].
\end{eqnarray}
Here $E_{\pm}=E_{\mathbf{p}\pm\mathbf{q}/2}$, $f_{\pm}=f(E_{\pm})$,
and $\omega$ implicitly has a small imaginary part. The factor ``sgn''
is $1$ and $-1$ in the density and spin channels respectively.  Note
that as a result of these sign changes in Eq.~(\ref{Chif}), the
dynamical correlations in the spin channel depend only on the square
of the pairing gap $\Delta$ which involves the sum of the squares of
$\Delta_{pg}$ and $\Delta_{sc}$. This is very different from the
particle density channel which reflects the distinction between
condensed (sc) and non-condensed (pg) pair contributions.

We now establish that the sum rules (Eq.(\ref{fs1})) are satisfied
along with Eq.~(\ref{fs2}). For illustration purposes, we choose a
fixed momentum transfer $q=0.5k_F$, and list the results of our sum
rule checks for various scattering lengths $k_Fa$ and two different
temperatures (above and below $T_c$) in Table~\ref{tab1}. 
Both sum
rules are satisfied to within $10\%$ or better. In the table
$\omega_M$ is the maximal frequency up to which we integrate in order
to satisfy the sum rules. We found that both structure factors approach
zero. 
On the
BCS side, the structure factors decay extremely rapidly so that
$\omega_M$ need not be large, but on the BEC side of resonance, they
exhibit a long and slowly decaying tail.
The characteristic size of this upper bound is of interest
experimentally in measurements of the structure factor (which reflect
the frequency integrated form of $S(\omega, \mathbf{q})$) such as in
Ref.~\onlinecite{Drummond10}.  We have also verified that the $f$-sum
rule holds for large momentum transfers; at unitarity
where our studies are most complete, the errors are smaller than
$5\%$.

Fig.~\ref{fig:StaticS} presents a plot of the the static structure
factors $S_c({\mathbf{q}})/2$ and $S_-({\mathbf{q}})$ as a function the
momentum transfer at unitarity and for two different temperatures
below and above $T_c$. The large $q$ behavior at the lower $T$ is
consistent with the $T=0$ results in Ref.~\onlinecite{Combescot06}.
At small $q$, all structure factors approach zero as required by
Eq.(\ref{fs2}). Fig.\ref{fig:StaticS}b is a central figure of this
paper. Here we plot $S_-({\mathbf{q}})$ as a function of temperature,
which one can see behaves like an order parameter. The same result
can be obtained for any fixed $\mathbf{q}$ and $\omega$ and reflects the
behavior implicit in Eq.~(\ref{Spm}). \textit{That is, the difference
between the density and spin channels arises only in the presence of a
condensate}.  This behavior is also found in strict BCS theory
as well as for any $k_F a$.  The vanishing of $S_-$ signals the vanishing of
condensation. It should establish if an unknown Fermi gas is in the
superfluid or normal phase.

Fig.~\ref{fig:DDS} shows the particle density (solid lines) and spin
response (dashed lines) in units of $n_F/2\epsilon_F$ as a function of
$\omega/\epsilon_F$. We show temperatures both below and above $T_c$
for a fixed momentum transfer $q=0.5k_F$ and for various $k_Fa$. Above
$T_c$, the magnetic and density response contributions coincide.
Below $T_c$, one sees from Fig.~\ref{fig:DDS} that $S_c(\omega,
\mathbf{q})$ has two main features, a collective mode peak, and a
continuum of quasiparticle excitations which appears for
$\omega>2\Delta$, associated with the breaking of Cooper pairs induced
by the perturbation.  The ``collective'' peak is broadened somewhat
due to coupling to the continuum. As the system evolves from the BCS
to the BEC regime (where the pairs are more difficult to break), this
``collective'' peak contains increasingly more of the total spectral
weight contained in the $f$-sum rule.  Above $T_c$, the spectral
weight in the former ``collective'' peak must be redistributed,
leading to an increased contribution in the continuum and to a new low
frequency peak which is associated with thermally broken pairs. This
peak appears at unitarity and on the BCS side of resonance for the
considered range of temperatures. At a more formal level it arises
from the contributions from the AL diagrams when
$\mu>\frac{1}{2m}(\frac{q}{2})^2$ above $T_c$.  These two continua
nearly merge in the BCS regime in Fig.~\ref{fig:DDS}(a); they are 
separated for the unitary and BEC cases.

Turning now to the spin structure factor at low $T$ (denoted by dashed
lines in Fig.~\ref{fig:DDS}), it can be seen that there is no
collective physics. The entire spectral weight of the $f$-sum rule is
associated with the continuum, and this, in turn, only appears when
$\omega>2\Delta$ where the energy transfer is large enough to break
pairs.  Just as for the particle density response, we see that the low
frequency spectral weight is highly suppressed.  In the BEC regime, in
Fig.~\ref{fig:DDS}(d), the spectral weight does not change significantly
with $T$, since the temperatures considered were not sufficient to
break pairs. This behavior is consistent with that found earlier
\cite{BaymPRA06} on the BEC side of resonance.  There is also a small
low frequency peak, which gives a tiny contribution to the $f$-sum
rule.

In this paper we proposed a 
methodology for establishing phase coherence
in neutral superfluids. This has been a long standing issue with one of the
earliest such proposals based on noise correlations \cite{Altman},
although this has proved difficult to implement away from $T=0$.
Here, by contrast we propose
two photon Bragg experiments at $T \neq 0$ which measure the difference
between the dynamical responses associated with the particle and spin density,
called 
$S_{-}(\omega,\mathbf{q})$ (or \cite{Combescot06,Drummond10} $S_{\uparrow \downarrow}
(\omega,\mathbf{q})/2$).
\textit{The qualitative behavior
shown in Figure
~\ref{fig:StaticS}(b), 
also obtains for 
$|S_{-}(\omega,\mathbf{q})|$
for all $\mathbf{q}, \omega$}.
This signature of phase coherent order, unlike others in the past
based on Bragg scattering, requires looking at a \textit{single} $\mathbf{q}, \omega, T$.
It should also apply to other crossover schemes which
satisfy the conservation laws, to trapped gases and to
optical lattices (at least) for frequencies below those of the
band gaps.

This work was supported by Grant No. NSF-MRSEC
DMR-0213745. 
C.C.C. acknowledges the support of the U.S.
Department of Energy through the LANL/LDRD Program.

\vspace*{-1ex}
\bibliographystyle{apsrev}

\begin{thebibliography}{18}
\bibitem[{wor({\natexlab{b}})}]{wordscollect}
\bibinfo{note}{We drop pairing correlations in the collective mode
  contribution, which leads to better agreement with the f-sum rule.}
\end{thebibliography}

\end{document}